# Erratum: Re-evaluation of neutron-$^4$He elastic scattering data near 20 MeV [Phys. Rev. C 83, 064616 (2011)]

M. Drosg, R. Avalos Ortiz, and B. Hoop

Correction for finite angular resolution in measurements of elastic scattering of neutrons from $^3$He was found to be significant [1]. Therefore, the prematurely published data of re-analyzed measurements of elastic scattering of neutrons from $^4$He are corrected for finite angular resolution. In the original experiment in question [2], a close geometry was chosen to result in high neutron flux incident upon the sample. That is, the distance of neutron source to center of the $^4$He scatterer was 11.5 cm, resulting in a mean incoming angular spread of ±6.6°. The opening angle of detected neutrons was 1.0°. Because neutron yield as a function of angle is not linear, the effect of finite angular spread is a shift from geometric mean of the scattering angle. The correct angle can be obtained by weighting the yield distributions over the subtended solid angles. Incoming and outgoing angular resolution corrections are dealt with separately. Instead of shifting the angle, a correction of the yield at the mean angle was applied. These corrections were determined by simulations under very small acceptance angles and also under actual experimental geometries. The original modeling of the experiment by the Monte Carlo code accounted just for the angle dependence of the *intensity* of the incoming neutrons over the acceptance angle but not for the finite angular resolution resulting from the incoming angular spread. Besides, this time the correct incoming energy of 17.71 MeV is used throughout rather than 17.72 MeV as done in the original paper.

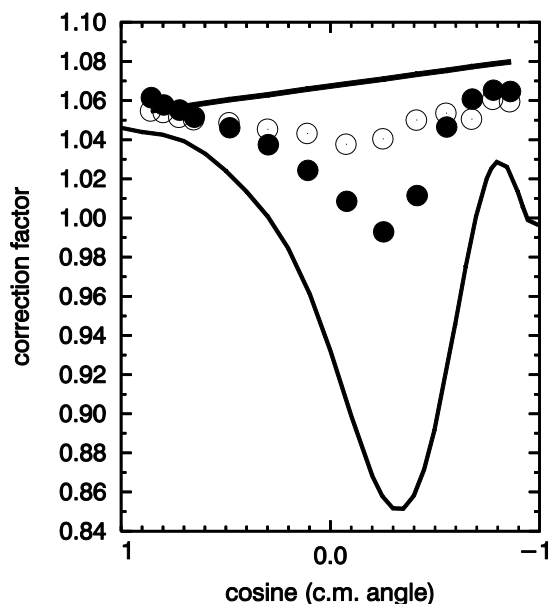

FIG. 1. 17.71 MeV neutron elastic scattering from helium. Comparison of the angular dependence of the original multiple scattering correction [2] (curve) with the presently used one (open dots). The full dots give the total correction (including the angular resolution correction). The simulation uncertainty is smaller than the size of the dots. The upper curve gives the correction values used previously.

TABLE I. Differential cross sections for elastic neutron scattering from $^4$He (angles in degrees, cross sections and uncertainties in mb/sr)

| | $E_n$=17.71±0.06 MeV | | | $E_n$=20.97±0.05 MeV | | | $E_n$=23.72±0.04 MeV | | |
|---|---|---|---|---|---|---|---|---|---|
| $\Theta_{lab}$ | $\Theta_{CM}$ | $\sigma_{CM}$ | $\Delta\sigma_{CM}$ | $\Theta_{CM}$ | $\sigma_{CM}$ | $\Delta\sigma_{CM}$ | $\Theta_{CM}$ | $\sigma_{CM}$ | $\Delta\sigma_{CM}$ |
| 20.0 | | | | 25.03 | 217.5 | 5.7 | 25.05 | 207.8 | 4.8 |
| 25.0 | 31.20 | 216.1 | 5.8 | 31.22 | 179.0 | 7.0 | | | |
| 30.0 | 37.35 | 188.7 | 5.8 | 37.66 | 158.7 | 4.0 | 37.38 | 150.5 | 3.2 |
| 35.0 | 43.43 | 163.1 | 4.3 | 43.45 | 139.1 | 3.1 | | | |
| 40.0 | 49.46 | 144.4 | 3.2 | 49.48 | 119.2 | 2.8 | 49.50 | 105.8 | 2.1 |
| 45.0 | | | | 55.44 | 103.9 | 2.4 | 55.46 | 82.3 | 1.7 |
| 50.0 | 61.29 | 101.1 | 3.0 | 61.31 | 80.4 | 1.4 | 61.34 | 67.8 | 1.7 |
| 55.0 | | | | 67.11 | 70.0 | 1.4 | 67.13 | 48.0 | 2.4 |
| 60.0 | 72.77 | 70.0 | 2.2 | 72.80 | 54.4 | 1.2 | 72.83 | 42.1 | 1.2 |
| 65.0 | | | | 78.41 | 43.13 | 0.93 | | | |
| 70.0 | 83.87 | 38.90 | 1.36 | 83.91 | 32.50 | 0.57 | 83.93 | 22.80 | 0.82 |
| 80.0 | 94.55 | 20.44 | 0.80 | 94.58 | 16.21 | 0.41 | 94.60 | 12.49 | 0.85 |
| 90.0 | 104.77 | 11.01 | 0.58 | 104.80 | 7.80 | 0.25 | 104.82 | 8.55 | 0.50 |
| 100.0 | 114.53 | 8.42 | 0.77 | 114.56 | 5.26 | 0.26 | 114.58 | 7.33 | 0.52 |
| 110.0 | 123.85 | 10.47 | 0.44 | 123.87 | 8.07 | 0.28 | 123.89 | 9.48 | 0.59 |
| 120.0 | 132.74 | 16.97 | 0.65 | 132.76 | 12.45 | 0.32 | 132.78 | 12.37 | 0.63 |
| 130.0 | 141.24 | 25.73 | 1.96 | 141.26 | 18.77 | 0.40 | | | |
| 140.0 | 149.41 | 30.15 | 0.80 | 149.43 | 25.59 | 0.44 | | | |

Two sets of simulated neutron distributions from neutron elastic scattering on $^4$He at 17.71, 20.97, and 23.72 MeV were obtained with a narrow beam and with actual beam conditions using the Monte Carlo neutron transport code MCNPX [3]. Because the data base of MCNPX for $^4$He(n,n)$^4$He covers only energies up to 20 MeV, it was extended to 23.8 MeV using the LANL1971 data (version 1978) [4]. The target used in the simulation was liquid Helium-4 (density 0.1294 g/cm$^3$), within stainless steel inner and outer container walls of combined thickness 0.018 cm, and ambient atmospheric pressure (0.96 mg/cm$^3$). The kinematics and geometric situation of the neutron source was simulated by a pencil source of 3.0 cm length and (energy dependent) intensity distribution of the $^3$H(d,n)$^4$He source into the forward direction. Actual geometric measures were used in the simulation.

Yield values were corrected by factors which account for the shift from the actual (weighted) mean angle to the nominal angle. This also corrects the shallower angular distribution minima. The results of the Monte Carlo simulation were checked numerically at several angles. Correction of yield values due to actual angular resolution was as large as 6.2% (at 100° in 20.97-MeV distribution) for the ingoing part, and less than 0.4% for the outgoing part. The latter was disregarded.

Figure 1 shows the corrections applied to the 17.71 MeV experimental angular distribution of $^4$He(n,n)$^4$He. The curve represents the original, exaggerated sample-size correction [2], the open dots the improved sample-size correction, and the full dots the present (sample-size + angular resolution) corrections. As discussed in Sect. V of the original paper the multiple scattering correction was not applied to the data published then, under the assumption that these contributions had been subtracted during the raw data analysis resulting in data that are possibly high by 1.5%. In this erratum we have included the simulated multiple scattering correction (open dots in Fig. 1) arriving at total corrections as shown in this Fig. for 17.71 MeV by full dots. The corrected data are given in Table I. The angular resolution correction contributes an estimated uncertainty of ±0.1% to the scale uncertainties, and ±<0.4% to the shape uncertainties. When added quadratically, there is no noticeable increase in the combined uncertainty.

TABLE II. Results of weighted least squares fitting using Legendre coefficients. Neutron energy $E_n$ in MeV, $\sigma_0$ in mb/sr, $\sigma_{el}$ in mb.

| $E_n$ | $\sigma_0$ | $A_0$ | $A_1$ | $A_2$ | $A_3$ | $A_4$ | $A_5$ | $A_6$ | $\sigma_{el}$ | $\sigma_{el}$(1971) |
|---|---|---|---|---|---|---|---|---|---|---|
| 17.71 | 284 | 0.246 | 0.370 | 0.307 | 0.037 | 0.020 | 0.016 | 0.004 | 878 | 858 |
| 20.97 | 276 | 0.215 | 0.334 | 0.297 | 0.057 | 0.052 | 0.030 | 0.016 | 746 | 732 |
| 23.72 | 256 | 0.203 | 0.330 | 0.306 | 0.103 | 0.042 | 0.013 | 0.003 | 653 | 654 |

Table II presents Legendre coefficients of the absolute differential cross sections of the reaction $^4$He(n,n)$^4$He for the three neutron energies. Results in the center-of-mass system are summarized by Legendre polynomial expansions in the form

$$\frac{d\sigma(E,\Theta)}{d\Omega} = \frac{d\sigma(E,0°)}{d\Omega} \cdot \sum_i A_i P_i(\cos\Theta) = \sigma_0 \cdot \sum_i A_i P_i(\cos\Theta)$$

The present data coincide within uncertainty limits with total cross section data [5] (taking 32.9 mb [6] for the non-elastic cross section at 23.72 MeV). For each of the three neutron energies, the elastic cross sections and uncertainties are, respectively, $\sigma_{el}$ = 878±15, 746±14, and 653±14 mb. The common (correlated) scale uncertainty of 1.6% is not included.

1. M. Drosg, R. Avalos Ortiz, and P. W. Lisowski, Neutron Interactions with $^3$He revisited. 1. Elastic Scattering Around and Beyond 10 MeV, *Nucl Sci. Eng.* **172**, 1 (2012).
2. A. Niiler, M. Drosg, J. C. Hopkins, J. D. Seagrave, and E. C. Kerr, *Phys. Rev. C* **4**, 36 (1971).
3. MCNPX, a general-purpose Monte Carlo N-Particle eXtended radiation transport code, Los Alamos National Laboratory; https://mcnpx.lanl.gov/
4. M. Drosg, Los Alamos Scientific Laboratory, Report No. LA-7269-MS, 1978.
5. B. Haesner, W. Heeringa, H. O. Klages, H. Dobiasch, G. Schmalz, P. Schwarz, J. Wilczynski, and B. Zeitnitz, *Phys. Rev. C* **28**, 995 (1983).
6. M. Drosg, IAEA Report No. IAEA-NDS-87 Rev. 9, 2005.